# Quantum electromagnetic "black holes" in a strong magnetic field

Igor I. Smolyaninov

*Department of Electrical and Computer Engineering, University of Maryland, College Park, MD 20742, USA*

**As demonstrated by Chernodub, strong magnetic field forces vacuum to develop real condensates of electrically charged $\rho$ mesons, which form an anisotropic inhomogeneous superconducting state similar to Abrikosov vortex lattice. As far as electromagnetic field behaviour is concerned, this state of vacuum constitutes a hyperbolic metamaterial [1]. Here we demonstrate that spatial variations of magnetic field may lead to formation of electromagnetic "black holes" inside this metamaterial. Similar to real black holes, horizon area of the electromagnetic "black holes" is quantized in units of the effective "Planck scale" squared. The magnetic fields of the required strength and geometrical configuration may be created on Earth in heavy-ion collisions at the Large Hadron Collider. We evaluate electromagnetic field distribution around an electromagnetic "black hole" which may be created as a result of such collision.**

Recent observation that vacuum in a strong magnetic field behaves as a hyperbolic metamaterial [1] opens up quite a few interesting research areas at the intersection of metamaterial optics, high energy physics and gravitation theory. For example, a well known "additional wave" solution of macroscopic Maxwell equations describing metamaterial optics in the presence of dispersion has led to prediction of ~ 2 GeV "heavy photons" in vacuum subjected to a strong magnetic field [2]. On the other hand, many non-trivial spacetime metrics considered in gravitation theory, such as black holes [3-6], wormholes [7,8], spinning cosmic strings [9], and metric near the big bang [10] find interesting analogues in metamaterial optics. As far as hyperbolic metamaterial behaviour of vacuum in a strong magnetic field is concerned, the most relevant result in this respect is theoretical prediction of electromagnetic "black holes" in fluctuating hyperbolic metamaterials [6].

Recent examination of electromagnetic fields generated as a result of Pb+Pb heavy ion collisions at LHC demonstrate that magnetic fields of the order of $10^{16}$ Tesla are generated during off-central collisions [11]. As shown by Chernodub [12,13], magnetic fields of this magnitude force vacuum to develop real condensates of electrically charged $\rho$ mesons, which form an anisotropic inhomogeneous superconducting state similar to Abrikosov vortex lattice [14]. As far as electromagnetic field behaviour is concerned, this state of vacuum behaves as a hyperbolic metamaterial [1]. Here we demonstrate that similar to ref. [6], spatial variations of magnetic field during the off-central heavy ion collisions at LHC may lead to formation of the electromagnetic metamaterial "black holes". Moreover, similar to real black holes [15], surface area of the electromagnetic "black holes" is quantized in units of the effective "Planck scale" squared. The latter result is natural since the effective metamaterial "Planck scale" [2] defined by the spatial scale at which the hyperbolic metamaterial exhibits strong spatial dispersion appears to be defined by the periodicity of the Abrikosov vortex lattice $L_B = \sqrt{2\pi\hbar/|eB|}$ ~0.5 fm [12,13]. Since the horizon area of the electromagnetic "black hole" is quantized in terms of the area $L_B^2$ of the Abrikosov lattice unit cell, electromagnetic "black holes" replicate the well known area quantization result [15] obtained for their real gravitation theory counterparts.

Before proceeding to actual derivation of these results, let us briefly summarize the basic features of macroscopic electrodynamics of the hyperbolic state of vacuum in a strong magnetic field. As demonstrated by Chernodub [12,13], a strong magnetic field forces vacuum to develop superconducting condensates of electrically charged $\rho$ mesons, which form an anisotropic superconducting state similar to Abrikosov vortex lattice in a type-II superconductor [14]. This unusual effect follows from the consideration of the ground state energy of the spin=1 charged $\rho$ mesons in an external magnetic field $B$ [12,13]:

$$m_\rho^2(B) = m_\rho^2 - |eB|, \qquad (1)$$

which indicates that at large magnetic fields

$$B > B_c = \frac{m_\rho^2}{e} \approx 10^{16} T \qquad (2)$$

vacuum must spontaneously generate positively and negatively charged $\rho$ meson condensates. These condensates form a periodic Abrikosov-like lattice of superconducting vortices separated by gaps of the order of $L_B = \sqrt{2\pi\hbar/|eB|}$ ~0.5 fm (Fig.1). Since superconductivity of charged $\rho$ mesons is realized along the axis of magnetic field only, there is no conductivity perpendicular to magnetic field at $T=0$. Therefore, effective dielectric tensor of vacuum in a strong magnetic field is strongly anisotropic. The diagonal components of the tensor are positive $\varepsilon_x = \varepsilon_y = \varepsilon_1 > 0$ in the x and y directions perpendicular to the magnetic field, and negative $\varepsilon_z = \varepsilon_2 < 0$ in the z direction along the field. As a result, vacuum in a strong magnetic field behaves as a hyperbolic metamaterial medium [1]. While hyperbolic metamaterials exhibit strong temporal dispersion, their spatial dispersion is typically modest. Therefore, it is possible to write macroscopic Maxwell equations in the frequency domain around a given frequency $\omega_0$ as [1,2]:

$$\frac{\omega^2}{c^2}\vec{D}_\omega = \vec{\nabla} \times \vec{\nabla} \times \vec{E}_\omega \text{ and } \vec{D}_\omega = \vec{\vec{\varepsilon}}_\omega \vec{E}_\omega \qquad (3)$$

Spatial distribution of monochromatic extraordinary electromagnetic wave inside a hyperbolic metamaterial may be described by the following wave equation for $E_{\omega z} = \varphi_\omega$:

$$-\frac{\partial^2 \varphi_\omega}{\varepsilon_1 \partial z^2} + \frac{1}{(-\varepsilon_2)}\left(\frac{\partial^2 \varphi_\omega}{\partial x^2} + \frac{\partial^2 \varphi_\omega}{\partial y^2}\right) = \frac{\omega_0^2}{c^2}\varphi_\omega = \frac{m^{*2} c^2}{\hbar^2}\varphi_\omega \qquad (4)$$

This equation coincides with the 3D Klein-Gordon equation describing a massive scalar field $\varphi_\omega$ propagating in a flat (2+1) dimensional effective Minkowski spacetime, in which spatial coordinate $z$ behaves as a "timelike" variable.

In ref.[1] effects of spatial dispersion were neglected, and effective permittivities of the hyperbolic state of vacuum were calculated in the limit $B \approx B_c$ as

$$\varepsilon_2 \approx \alpha \varepsilon_s < 0, \quad \varepsilon_1 = \frac{1+\alpha}{1-\alpha} > 0 \qquad (5)$$

where $\varepsilon_s$ is the dielectric permittivity of the superconducting phase, and the average volume fraction of the superconducting phase $\alpha$ is small. The volume fraction of the superconducting phase can be estimated as [1]

$$\alpha \approx 4\pi r_\rho^3 \rho(B)/3m_\rho, \tag{6}$$

where $r_\rho \sim 0.25$fm is the $\rho$ meson radius, $m_\rho$ is the $\rho$ meson mass, and the $\rho$ meson condensate density is

$$\rho(B) = C_\phi \frac{m_q(B)}{G_V}\left(1 - \frac{B_c}{B}\right)^{1/2}, \quad \text{at } B > B_c \tag{7}$$

$$\rho(B) = 0 \quad \text{at } B < B_c$$

where $C_\phi = 0.51$ is a constant, $m_q(B)$ is the quark mass, and $G_V$ is the vector coupling of four-quark interactions [12,13]. Effects of spatial dispersion on the hyperbolic state of vacuum were evaluated in ref.[2]. In the limit of small volume fraction $\alpha$ of the superconducting "wires", $\varepsilon_1 \approx 1$ does not exhibit any spatial dispersion. On the other hand, $\varepsilon_2$ does exhibit considerable spatial dispersion due to plasma excitations of individual "wires" [2,16]:

$$\varepsilon_2 = 1 + \frac{1}{\dfrac{1}{\alpha(\varepsilon_s - 1)} - \dfrac{\omega^2/c^2 - k_z^2}{k_p^2}}, \tag{8}$$

where

$$(k_p L_B)^2 \approx \frac{2\pi}{\ln\left(\dfrac{L_B}{\pi d}\right) + 0.5275}, \tag{9}$$

and $d$ is the "wire" diameter. In the limit of small $\omega$

$$\varepsilon_2 \approx \alpha\varepsilon_s - \frac{k_z^2 \alpha^2 \varepsilon_s^2}{k_p^2}, \tag{10}$$

so that the "effective Planck scale" [2] may be introduced as

$$k_{pl} \approx \frac{k_p}{|\alpha\varepsilon_s|^{1/2}} \qquad (11)$$

Spatial dispersion at this scale cannot be neglected, and the effective 3D "Lorentz symmetry" of eq.(4) is broken at this effective "Planck scale".

Let us demonstrate that similar to regular hyperbolic metamaterials [6], spatial variations of $\varepsilon_2$ (due to spatial variation of magnetic field $B$) may lead to appearance of electromagnetic "black holes" in the hyperbolic state of vacuum in a strong magnetic field. We begin with the Klein-Gordon equation for a massive particle in a gravitational field [17]

$$\frac{1}{\sqrt{-g}}\frac{\partial}{\partial x^i}\left(g^{ik}\sqrt{-g}\frac{\partial \varphi}{\partial x^k}\right) = \frac{m^2c^2}{\hbar^2}\varphi \qquad (12)$$

Our goal is to emulate particle behavior in the Rindler metric:

$$ds^2 = -g^2z^2dt^2 + dx^2 + dy^2 + dz^2, \qquad (13)$$

which has a horizon at $z=0$. The resulting Klein-Gordon equation is

$$-\frac{1}{g^2z^2}\frac{\partial^2\varphi}{\partial t^2} + \frac{\partial^2\varphi}{\partial x^2} + \frac{\partial^2\varphi}{\partial y^2} + \frac{\partial^2\varphi}{\partial z^2} + \frac{1}{z}\frac{\partial\varphi}{\partial z} = \frac{m^2c^2}{\hbar^2}\varphi \qquad (14)$$

By changing variables to $\psi = z^{1/2}\varphi$ and $\tau = gzt$, the latter equation may be rewritten as

$$-\frac{\partial^2\psi}{\partial\tau^2} = \left(-\frac{\partial^2}{\partial x^2} - \frac{\partial^2}{\partial y^2} - \frac{\partial^2}{\partial z^2} - \frac{1}{4z^2} + \frac{m^2c^2}{\hbar^2}\right)\psi \qquad (15)$$

Eq.(15) demonstrates that an electromagnetic "black hole" may be emulated if we manage to reproduce approximately the following dispersion relation inside a metamaterial:

$$\frac{\omega^2}{c^2} = k_\perp^2 + k_z^2 - \frac{1}{4z^2} + \frac{m^2 c^2}{\hbar^2} \qquad (16)$$

Thus, near the horizon in the limit $z \to 0$ we must produce $k_z \sim z^{-1}$. The dispersion relation of extraordinary photons inside the hyperbolic metamaterial vacuum state given by eq.(4) is

$$\frac{\omega^2}{c^2} = \frac{k_\perp^2}{\varepsilon_2} + \frac{k_z^2}{\varepsilon_1} \approx \frac{k_\perp^2}{\alpha \varepsilon_s} + k_z^2 \qquad (17)$$

where $\varepsilon_s$ is negative. Therefore, a spatial distribution of $\alpha$ which may be approximated as $\alpha \sim z^2$ would do the job. Such a distribution would emulate an event horizon near $z=0$. Taking into account eqs.(6,7), we obtain the desired spatial dependence of magnetic field inside the hyperbolic metamaterial state of vacuum:

$$B = B_c \left(1 + \gamma z^4 \right), \qquad (18)$$

where $\gamma$ is a positive constant. Such a spatial distribution of magnetic field will be able to emulate an electromagnetic "black hole" for a considerable range of $k_\perp \neq 0$. On the other hand, as evident from eq.(17) this approximation fails for $k_\perp \approx 0$. However, this limitation is not meaningful due to experimental constraints: it is not feasible to produce a spatially homogeneous superconducting/metamaterial state of vacuum in a large enough spatial volume so that behavior of $k_\perp \approx 0$ electromagnetic modes would be observable.

Let us demonstrate that spatial distribution of magnetic field described by eq.(18) arises naturally as a result of off-central heavy ion collisions at LHC. Typical geometry of such a collision just before the impact is illustrated in Fig.2. Due to relativistic velocities of the ions, they are Lorentz contracted to pancake shapes.

According to recent theoretical calculations [11], magnetic field just before the collision is given by

$$B_\perp(0,\vec{r}) \approx \frac{es^{1/2}}{8\pi m_n c^2} \sum_n \frac{\vec{e}_{ny} \times \vec{R}_{n\perp}}{R_{n\perp}^3}, \qquad (19)$$

which in the non-relativistic limit would reduce to the usual Bio-Savart's law for a set of moving charges. Here we assume $s^{1/2}$=3.8 TeV as a typical center-of-mass energy per nucleon pair during Pb + Pb collisions at LHC. Summation is performed over all the protons participating in the collision, $m_n$ and $\vec{r}_n$ are the mass and position of the n-th proton, $\vec{R}_n = \vec{r} - \vec{r}_n$, and $\vec{e}_{ny}$ is the unit vector in the $\pm y$ direction depending on whether the n-th proton belongs to the target or projectile ion. The resulting magnetic field $B_z$ distribution along the z-axis at the moment t=0 just before the collision is presented in Fig.3. Note that the contributions of the projectile protons located in the overlap region in Fig.2 mutually compensate each other. This effect results in two symmetric maxima of $B_z$ near the points marked as A and B in Fig.2, and the central minimum at z=0. As a result, magnetic field distribution near z=0 matches eq.(18), which describes a magnetic field distribution leading to appearance of the electromagnetic "black hole" event horizon at z=0. Moreover, magnitude of the magnetic field produced during the collision matches eq.(2).

We have performed numerical simulations of the AC extraordinary field distribution in the vicinity of the electromagnetic "event horizon" located at z=0 using COMSOL Multiphysics 4.2 solver. In these simulations the dipole sources of radiation are placed in the overlap region shown in Fig. 2 (which is natural since radiation is created during the heavy ion collision). Our results are presented in Fig.4. As expected, strong field enhancement is observed near z=0 and somewhat weaker enhancement is observed around points marked as A and B in Fig.2. Near these points $\alpha=0$ (or $B=B_c$) conditions are also realized in the hyperbolic metamaterial state of vacuum. However, field distribution around these secondary maxima does not satisfy eq.(18). The characteristic features of electromagnetic field distribution, which emerge from our macroscopic electrodynamics analysis, appear to be tightly localized on the femtometer scale (see Fig.4), which corresponds to $3 \times 10^{-24}$ s time scale (this is much shorter than the collision time scale $\tau \sim 2d/c \sim 10^{-22}$ s). Thus, the frequency domain approach to macroscopic Maxwell equations appears to be well justified. We should point out that the degree of field divergence at the "horizon" in these simulations is defined by the metamaterial losses, which is not well known. Further improvement of this model would need to take into account such parameters as the equation of state of vacuum in a strong magnetic field and nonzero temperature after the collision. A quark-gluon plasma created during the collision should have a very high temperature which may destroy the superconductive/metamaterial state of vacuum. On the other hand, this temperature-related complications may be avoided in an ultra-peripheral collision in which the impact parameter $b \geq d$, where $d$ is the diameter of almost-colliding nuclei/ions. Since the nuclei do not touch in an ultra-peripheral collision, the quark-gluon plasma is not created and the temperature is close to zero. On the other hand, the magnetic field created as a result of such ultra-peripheral collision will have about the same strength.

Short time of the collision $\tau \sim 2d/c \sim 10^{-22}$ s may also be a problem for the proposed effect. Such a short collision time may not be sufficient to create the superconducting metamaterial state of vacuum. While $\tau$ is one order of magnitude longer than the characteristic time scale $\tau_s \sim 10^{-23}$ s of strong interactions, the superconducting/metamaterial ground state is a new state, and vacuum needs certain transition time to pass from the standard (insulator) phase to the new metamaterial phase. Thus, even though $\tau >> \tau_s$, there is a chance that this transition may not be completed. This issue will require an additional effort, which is beyond the scope of the current paper.

We must also note that according to Fig.3 the spatial extent of large magnetic fields $B>B_c$ upon collision is considerably larger than $L_B$, which means that the macroscopic electrodynamics description of the system is applicable. On the other hand, the effective "horizon" area at $z=0$ is clearly quantized in terms of the area $L_B^2$ of the Abrikosov lattice unit cells, and only a few tens of unit cells would fit into the horizon area. Therefore, the electromagnetic "black holes" created during the off-central heavy ion collisions at LHC must be considered "quantum". They must replicate the well known area quantization result [15] obtained for the real black holes.

As demonstrated in ref.[2], creation of the superconducting/metamaterial state during the heavy-ion collisions at LHC may manifest itself via excess of particle jets with ~2 GeV energy, which corresponds to creation of "heavy photon" states during the collision. The origin of these "heavy photon" states is similar to appearance of the extraordinary "additional wave" in the wire array hyperbolic metamaterials [18]. The heavy photons are TM polarized, which means that their magnetic field is normal to the axis of the external applied magnetic field (the z-direction). Such heavy photons can propagate in any direction inside the hyperbolic metamaterial state [2]. If on the other hand, a quantum electromagnetic "black hole" is created within the superconducting/metamaterial state, it will affect angular distribution of these heavy photons, and hence angular distribution of the ~2 GeV particle jets. Attractive effective potential of the electromagnetic "black hole" will lead to particle jets concentration towards the *(x,y)* plane.

In conclusion, we have demonstrated that spatial variations of magnetic field may lead to formation of electromagnetic "black holes" inside the hyperbolic metamaterial state of vacuum in a strong magnetic field. Similar to real black holes, horizon area of the electromagnetic "black holes" is quantized in units of the effective "Planck scale" squared. The magnetic fields of the required strength and geometrical configuration may be created on Earth in heavy-ion collisions at the Large Hadron Collider. We have evaluated electromagnetic field distribution around an electromagnetic "black hole" which may be created as a result of such collision.

**Figure Captions**

**Figure 1.** Abrikosov lattice of charged ρ meson condensates forming a "wire array" hyperbolic metamaterial structure of vacuum in a strong magnetic field [1]. The pictorial diagram is drawn schematically to provide visual representation of anisotropic character of superconductivity of the $\rho$ meson condensate: it superconducts only in the direction along magnetic field, while conductivity in the perpendicular direction is zero at T=0.

**Figure 2.** Geometry of the off-central heavy ion collision with an impact parameter *b*. Circles represent the colliding ions with diameter *d*. The ions are Lorentz contracted to pancake shapes and move in the opposite directions along the *y* axis.

**Figure 3**. (a,b) Calculated magnetic field $B_z$ distribution along the *z*-axis at the moment *t*=0 just before the collision is shown at different scales. Field distribution near *z*=0 matches eq.(18), which describes a magnetic field distribution leading to appearance of the electromagnetic "black hole" event horizon at *z*=0.

**Figure 4**. Numerical simulations of the extraordinary electromagnetic field distribution in the vicinity of the electromagnetic "event horizon" located at *z*=0.

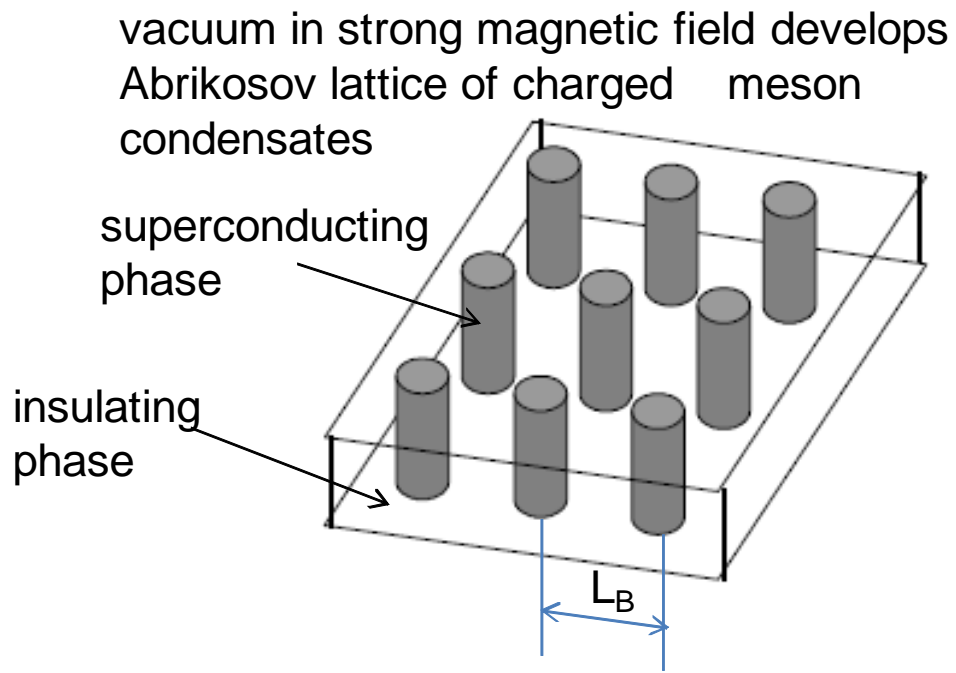

Fig.1

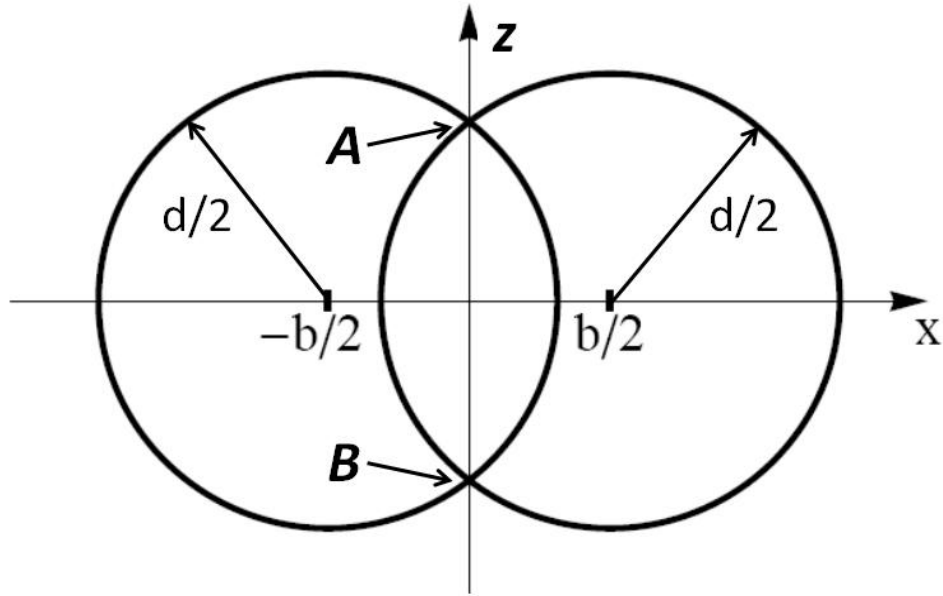

Fig. 2

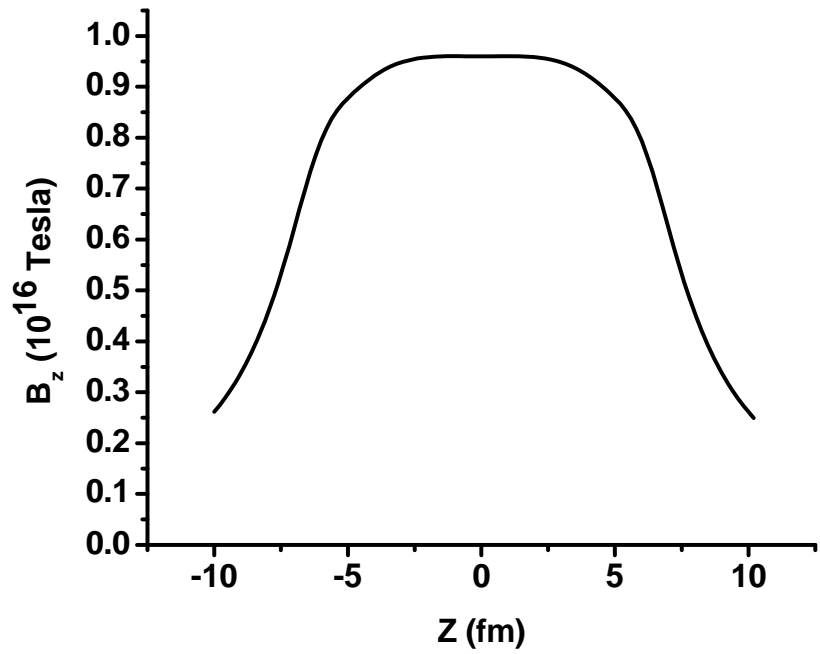

(a)

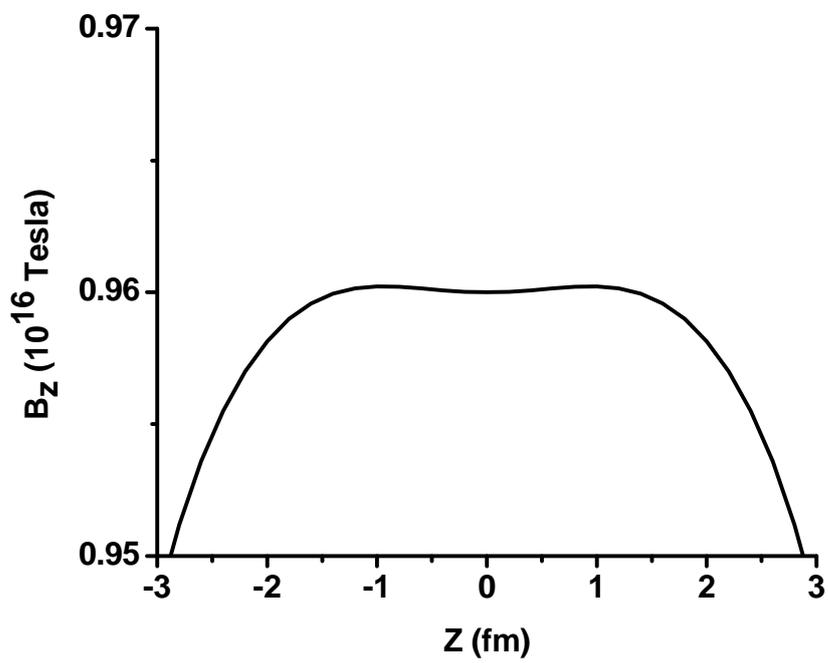

(b)

Fig. 3

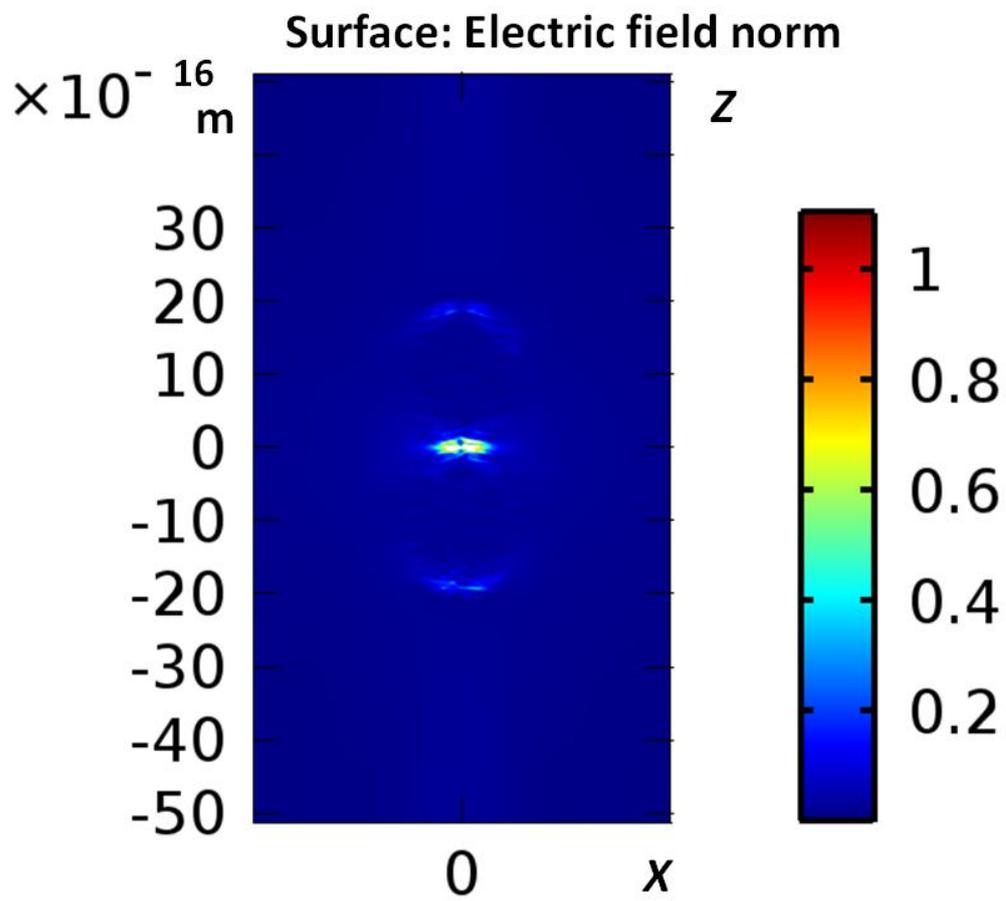

Fig. 4